\documentclass[aps,pre,superscriptaddress,singlecolumn]{revtex4}

\usepackage{amssymb}
\usepackage{graphicx}
\usepackage{colordvi}
\usepackage[dvipsnames]{xcolor}
\usepackage{braket}
\usepackage{bold-extra}
\usepackage{dsfont}
\usepackage{amssymb}
\usepackage[T1]{fontenc}

\usepackage{subfigure}

\usepackage{amsmath}

\usepackage[utf8]{inputenc}

\graphicspath{{./figures}}

\begin{document}

\title{Exact quantum dynamics of Fermi--Hubbard systems using the Gaussian phase-space representation with diffusion gauges} %

\author{François Rousse}
\affiliation{Department of Physics and Astronomy, Uppsala University, Uppsala, Sweden}
\author{Massimiliano Fasi}
\affiliation{Department of Computer Science, Durham University, Stockton Road, DH1 3LE, UK}
\author{Andrii Dmytryshyn}
\affiliation{Department of Mathematical Sciences, Chalmers University of Technology and University of Gothenburg, 41296 Gothenburg, Sweden}
\affiliation{School of Science and Technology, Örebro University, 70182 Örebro, Sweden}
\author{Mårten Gulliksson}
\affiliation{School of Science and Technology, Örebro University, 70182 Örebro, Sweden}
\author{Joel F. Corney}
\affiliation{School of Mathematics and Physics, University of Queensland, Brisbane, Queensland 4072, Australia}
\author{Magnus Ögren}
\affiliation{School of Science and Technology, Örebro University, 70182 Örebro, Sweden}
\affiliation{HMU Research Center, Institute of Emerging Technologies, 71004, Heraklion, Greece}

\date{\today}

\begin{abstract}

We use the Gaussian Phase-Space Representation to solve the real-time dynamic of interacting fermions in 1D, 2D, and 3D systems. The method is exact up to a spiking point, which represents a limit on the practical simulation time. The spiking can be delayed, and the practical simulation time extended, by adjusting the gauges of the representation, resulting in different equivalent stochastic differential equations. Here, we work on the so-called diffusion gauge and propose an algorithm to find efficiently new implementations of the noise terms. Compared with our initial results 
[F. Rousse \textit{et al.} 2024, J. Phys. A: Math. Theor. \textbf{57}, 015303],
the new method achieves a significantly longer practical simulation time and can be applied to significantly larger systems.

\end{abstract}

\newenvironment{mf}{\begin{quote}\sffamily\textbf{\color{blue}Max: }}{\end{quote}}

\maketitle

\section{Introduction} \label{sec:Introduction}

Because of quantum entanglement, representing a quantum many-body system in its full complexity requires encoding all correlations between particle states. The amount of information required grows exponentially with the number of particles, which restricts current quantum simulations to relatively small systems.  
A range of methods have thus been developed to simplify quantum calculations, by either reducing the Hilbert space dimension, or removing or simplifying  the interactions between particles. 
In the Density-Matrix Renormalization Group (DMRG) method~\cite{schollwock2005density}, for example, which has been very successful for 1D systems, the correlations that do not contribute to the target state are discarded. More recently, AI-driven methods~\cite{carleo2017solving} learn which correlations are important enough to be kept.
Instead of removing dimensions, phase-space representation methods map the wave function to a density probability in a phase space, whose dynamics can then be sampled stochastically to calculate the desired correlations as a function of time.

The idea of using phase-space methods for quantum systems emerged early with the Wigner quasiprobability ~\cite{wigner1932quantum}; later, other representations were developed~\cite{husimi1940some, glauber1963coherent} in which different operator orderings for the phase-space variable mappings were tried. These methods were first used for bosonic systems \cite{drummond1980generalised, drummond1987quantum, polkovnikov2010phase, wuster2017quantum} and later adapted to fermionic systems \cite{corney2005gaussian, corney2006gaussian, corboz2013phase, davidson2017semiclassical, rousse2023correlated} and systems that feature a mixture of bosons and fermions \cite{ogren2010first, ogren2011stochastic}.

Here, we focus on the Gaussian Phase-Space Representation (GPSR) \cite{corney2005gaussian, corney2006gaussian} applied to a Fermi--Hubbard model. 
The Fermi--Hubbard model remains relevant in materials science: it uses a simple framework to capture the essential physics of strongly correlated electron systems, such as magnetism \cite{greif2015formation} and superconductivity~\cite{aimi2007gaussian}, and it provides insights into the behavior of electrons in complex materials, including high-temperature superconductors and transition-metal oxides \cite{hubbard1963electron, gutzwiller1963effect, kanamori1963electron, Will2013}.

The GPSR applied to the Hubbard model generates a Fokker--Planck equation for the phase-space density function, with one-body operators in the Hamiltonian generating only drift terms and two-body terms leading also to diffusion terms. 
That means that we can sample the probability density by a cloud of points, each with an independent trajectory that follows a corresponding stochastic differential equation (SDE).

In the absence of boundary terms~\footnote{In deriving the Fokker-Planck equation, the assumption is made that the distribution vanishes sufficiently quickly at large radius, allowing certain boundary terms to be neglected.}, the Fokker-Planck equation generated by the GPSR is an exact representation of the evolving quantum state~\cite{corney2005gaussian}. A stochastic sampling of the dynamics via the SDEs, with an associated sampling error, leads to a finite precision, which can be statistically improved in principle by use of larger sets of trajectories until the desired precision is reached.
However, after a certain simulation time, boundary corrections may appear, or individual trajectories may tend to infinity, causing the numerical result to become unreliable from that point onward.
We refer to the time elapsed before the occurrence of these issues as the \textit{practical simulation time}.
When using imaginary time to represent the inverse temperature~$\sim T^{-1}$, similar boundary corrections hamper the calculations of fermionic ground-states~\cite{assaad2005symmetry, corboz2008systematic}, in the limit $T \rightarrow 0$.

Several authors have shown that these problems can be alleviated by means of projection methods~\cite{assaad2005symmetry, corboz2008systematic, aimi2007gaussian}.

For real-time evolution, the practical simulation time can be improved by exploiting the freedom in writing down the SDE noise terms (the so-called noise matrix) for a given form of the Fokker-Planck equation, called the diffusion gauge~\cite{plimak2001optimization}. Particular SDE noise matrices can be found analytically, as in Eq.~\eqref{eq:B0Matrix}, but they can also be created through singular-value decomposition (SVD) of the diffusion matrix. We call the noise matrices obtained using these two methods the ``analytical'' and ``numerical'' noise matrix or diffusion gauge, respectively.

Here, we propose a new method for diffusion gauges~\cite{rousse2023simulations} in the fermionic GPSR~\cite{corney2005gaussian,corney2006gaussian}.
We assess the new technique by benchmarking quantum dynamics of many-body systems against other methods.
We also illustrate the phenomenon of spiking trajectories in calculations of fermionic quantum dynamics.
In particular, we show by numerical examples that the use of numerical diffusion gauges can delay the onset of spiking trajectories, and hence prolong the practical simulation time.
A second advantage of numerical diffusion gauges is that they constitute a reliable procedure for determining noise matrices from a diffusion matrix, in contrast to analytical noise matrices which are typically found using a ``guess and check procedure'', although various specific examples are known~\cite{ogren2010first, ogren2011stochastic, corboz2013phase}.

In section \ref{sec:method}, we explain how we calculate the quantum dynamic with GPSR and introduce our approach, which is based on the randomized SVD, to calculate the noise matrix from the diffusion matrix as a gauge. In section \ref{sec:num-exp}, we compare this new method to existing alternatives, showing that it improves in terms of both practical simulation time and computational performance.
Finally, in section~\ref{sec:conclusion}, we list a number of directions in which this work could be taken further.

\section{Method}
\label{sec:method}

\subsection{Fermi--Hubbard Hamiltonian}
\label{sec:FHHam}

We consider the time dynamic of a Fermi--Hubbard system of $n_s$ sites, whose general Hamiltonian, written in the second quantization formalism, is
\begin{equation}
    \label{eq:FHeq1}
    \hat{H} = -  \sum_{  i,j , \sigma} J_{ij} \hat{c}_{i \sigma}^{\dagger} \hat{c}_{j \sigma} +  \sum_{i,\sigma_i,j,\sigma_j} U_{i\sigma_ij\sigma_j}  \hat{c}_{i \sigma_i}^{\dagger}  \hat{c}_{j \sigma_j}^{\dagger}   \hat{c}_{j \sigma_j} \hat{c}_{i \sigma_i},
\end{equation}
where $\hat{c}_{i \sigma}$ and $\hat{c}_{i \sigma}^{\dagger}$ are the annihilation and creation operators on site $i$, respectively, of a $\sigma$-spin particle $\sigma \in \{ \uparrow, \downarrow \}$, $J_{ij}$ is the tunneling interaction between sites $i$ and $j$, and $U_{ij}$ is the Coulomb interaction between two particles on sites $i$ and $j$. Here we choose the common case of constant tunneling interactions between nearest neighbors only ($J_{ij} = t \delta_{<ij>}$) and of constant on-site Coulomb interactions ($u_{ij} = u\delta_{ij}$). Because of the on-site Coulomb interaction and the Pauli exclusion, only particles of opposite spin interact. Hence, our Hamiltonian is
\begin{equation}
    \label{eq:FHeq0}
    \hat{H} = -  J \sum_{  <i,j> , \sigma}  \hat{c}_{i \sigma}^{\dagger} \hat{c}_{j \sigma} +  U \sum_{i}   \hat{c}_{i \uparrow}^{\dagger}  \hat{c}_{i \downarrow}^{\dagger}  \hat{c}_{i \downarrow} \hat{c}_{i \uparrow} .
\end{equation}
Although we focus on the case of onsite interactions, the methods presented below are also applicable to the more general classes of interactions. In the numerical examples below, we set $J=U=1$, corresponding to a challenging regime computationally since tunneling and interaction energies are of similar scale.

\subsection{Time-dependent dynamic}
\label{sec:time-dep-dyn}

In this work on quantum dynamics, we solve the time-dependent Schr\"{o}dinger equation
\begin{equation}
    \label{eq:ShrEq}
     \frac{d}{dt} \ket{\Psi} = -\frac{i}{\hbar} \hat{H} \ket{\Psi}.
\end{equation}

We compare the phase-space representation method with the corresponding mean-field approximation, the Hartree--Fock (HF)~\cite{rousse2023correlated} approximation, and, if the system is small enough, with the exact diagonalization (ED) method. For the ED solution, we write the wave-function $\Psi$ in its fermionic Fock states basis with constant number of particles, and we use the Krylov subspace projection technique \cite{sidje1998expokit} to update the wave function of a small time step $dt$
\begin{equation}
    \label{eq:ShrSol}
     \ket{\Psi(t+dt)} =  e^{ -\frac{i}{\hbar} \hat{H} dt} \ket{\Psi(t)}.
\end{equation}

Depending on the connectivity of the system and the amount of computer memory available, the ED method can model systems up to about 20 sites.

For the HF solution, we write the one-particle density matrix in its local basis $n_{ij\sigma} = \braket{\hat{c}_{j\sigma}^{\dagger} \hat{c}_{i\sigma}} $ and we use Wick's theorem to factorise two-body terms
\begin{equation}
\label{eq:WickTh}
\braket{\hat{c}^{\dagger}_{i}\hat{c}^{\dagger}_{j}\hat{c}_{k}\hat{c}_{l}} =n_{il}n_{jk} - n_{ik}n_{jl} ,
\end{equation}
which gives us a closed set of equations for the one-particle density matrix. 
We integrate these equations with a 4th order Runge--Kutta method. Comparing GPSR to both ED and the mean-field approximation will help us understand the relevance of the new many-body simulation method.

\subsection{Fermionic Gaussian phase-space representation}
\label{sec:ferm-gauss-phase}

In phase-space methods, the many-body system is described with a multi-dimensional probability density. Using an over-complete basis of operators $\hat{\Lambda}$, parameterized by $\vec{\lambda}$, we expand the density matrix
\begin{equation}
    \hat{\rho}= \int  P(\vec{\lambda}) \hat{\Lambda}(\vec{\lambda}) d \vec{\lambda} .
\end{equation}
Here, $P(\vec{\lambda})$ contains all the information of the density matrix, and the expectations of every observable $\hat{O}$ can be computed by integrating the density with the corresponding phase-space function $O(\vec{\lambda})$, which gives
\begin{equation}
   Tr \big[\hat{O}\hat{\rho} \big] = \langle \hat{O} \rangle = \int  P(\vec{\lambda}) O(\vec{\lambda}) d \vec{\lambda} = \langle O \rangle_{P} .
\end{equation}

The general form of the basis of operators, the link between an observable and its corresponding phase-space function, and the properties of the probability density $P$ are explained in details by Corney and Drummond~\cite{corney2005gaussian, corney2006gaussian}. In the fermionic Gaussian-representation, the operator basis, in its general form, are Gaussian forms of normally-ordered creation and annihilation operators, see \cite{corney2005gaussian, corney2006gaussian}. In our case, the Hamiltonian conserves the particle numbers, so the anomalous operators disappear, and we will not use the amplitude of the operators, $\Omega$ in \cite{corney2005gaussian}, for the simulations. The mapping between the observables and their corresponding phase-space function is quite direct: the number operators are the first-order moments of $P$, or $\langle \hat{n}_{ij\sigma} \rangle = \langle \hat{c}^{\dagger}_{i \sigma} \hat{c}_{j \sigma} \rangle = \langle n_{ij\sigma} \rangle_P$, and the number-number occupancy operators are combinations of second-order moments, or covariances, between variables of $P$. 

In order to establish the partial differential equation that governs the time evolution of $P$, we start from the master equation for $\hat{\rho}$ in the Heisenberg picture,
\begin{equation}
\label{eq:dHeis_rho}
     \frac{d}{d t} \hat{\rho}= -\frac{i}{\hbar} \left[ \hat{H}, \hat{\rho}\right] ,
\end{equation}
and apply on the right-hand-side of the equation the mapping between pairs of creation/annihilation operators on the density matrix $\hat{\rho}$ and linear density operators on the probability density $P(\vec{n})$ described in \cite{corney2006gaussian}. If we denote the Green's function for holes $\tilde{n}_{ij\sigma} = \delta_{ij} - n_{ij\sigma}$, then the two important relations are
\begin{equation}
\label{eq:mapping}
    \begin{split}
        \hat{c}^{\dagger}_{i\sigma} \hat{c}_{j\sigma} \hat{\rho} \longrightarrow  \left[ n_{ij\sigma} - \frac{\partial}{\partial n_{lk\sigma}} \tilde{n}_{ik \sigma} n_{lj \sigma} \right] P(\vec{n}) ,\\
        \hat{\rho} \hat{c}^{\dagger}_{i \sigma} \hat{c}_{j\sigma}  \longrightarrow  \left[ n_{ij\sigma} - \frac{\partial}{\partial n_{lk\sigma}} n_{ik\sigma}  \tilde{n}_{lj\sigma} \right] P(\vec{n})\\
    \end{split}
\end{equation}
which leads to the equation for the probability density being a Fokker--Planck equation
\begin{equation}
\label{eq:dPDE_n}
     \frac{\partial}{\partial t} P (\vec{n})= \left[ - \frac{\partial}{\partial n_{ij\sigma}} A_{ij\sigma}(\vec{n}) + \frac{1}{2} \frac{\partial^2}{\partial n_{ij\sigma} \partial n_{kl\sigma'}} D_{ij\sigma,kl \sigma'}(\vec{n})  \right] P(\vec{n}) .
\end{equation}

We give explicit forms for the drift vector $A_{ij\sigma}$ and the diffusion matrix $D_{ij\sigma,kl \sigma'}$ in the appendix \ref{app:FPMatrices}. We can recover the mean-field dynamic by truncating the Fokker--Planck equation at the first derivative order. We can also produce beyond-mean-field dynamics by imposing conditions on the probability density $P(\vec{n})$~\cite{rahav2009gaussian}. The full PDE (\ref{eq:dPDE_n}) is impractical to solve directly for more than a few modes, due to the large phase-space dimension.
However, we can write down and numerically solve the It{\^o} equations for the stochastic processes governed by this PDE: 

\begin{equation}
\label{eq:SDE_n}
\begin{split}
     d \vec{X}_t  &=  \vec{A}(\vec{X}_t) dt + B (\vec{X}_t) d \vec{W}_t \\
     \end{split} , 
\end{equation}
where $\vec{X}_t$ is a stochastic process whose dimensions are directly linked to those of $\vec{n}$, the noise matrix $B$ is defined from $BB^T = D$, and $\vec{W}_t$ is a vector of Wiener processes \cite{gardiner2009stochastic}. Hence we model the probability density by a set of trajectories $\vec{X}_t$ that follow their own independent trajectory directed by the stochastic differential equation (SDE) written in (\ref{eq:SDE_n}). The dimensions of the stochastic process $\vec{X}_t$ have a one-to-one correspondence to the PDE variables $\vec{n}$, the moments of $P(\vec{n},t)$ are recovered with discrete averages on the set of $M$ trajectories, for example $\langle n_{ij\sigma} \rangle_P=\sum_k X^k_t(i,j,\sigma)/M$, see \cite{gardiner2009stochastic}.

A true Fokker-Planck equation must have a positive-definite diffusion matrix when written in terms of real variables. The GPSR, like the positive-P method on which it is based, only specifies a Fokker-Planck equation in terms of complex variables. There is a freedom in how the derivatives with respect to complex variables can be written in terms of real derivatives.  It is this  freedom, which arises from the overcompleteness of the basis, which is exploited in the diffusion gauge.

\subsection{Diffusion Gauge}

A reason for GPSR to break down is that the probability density representing the density matrix broadens with time and ends up presenting fat tails \cite{gilchrist1997positive, drummond2003quantum}, leading to error in the sampling of observables. Practically, some trajectory will become unstable and escape towards infinity, which prevents GPSR from computing averages. This breakdown of the method is preceded by spikes in the observables: since only a finite number of trajectories is used, the large amplitudes reached by some trajectories cause the averages to spike before the simulation breaks down. If we cannot prevent this problem, we can try to delay it \cite{rousse2023simulations, perret2011stabilizing, plimak2001optimization}. Indeed, the basis is ``over-complete'', meaning that multiple probability densities $P$ can represent the same density matrix $\hat{\rho}$. Through choice of gauge we aim to engineer the distribution to remain more compact during time evolution. For each trajectory, the freedom in the choice of the SDEs is a ``gauge''. Here, we are interested in the so-called ``diffusion gauge'', which arises because the complex diffusion matrix admits multiple decompositions. 

We already know a decomposition of $D$, the so-called ``analytical'' decomposition $B_0$, described in appendix~\ref{app:FPMatrices}. The size of $B_0$ gives us an upper limit for the rank of the diffusion matrix. We use this rank to decompose efficiently the diffusion matrix into another noise matrix. Although using the analytical noise matrix directly is faster than computing another one, and hence allows the simulation of larger systems, the practical simulation time can be greatly improved by use of the new noise matrix~\cite{rousse2023simulations, plimak2001optimization, hitzelhammer2025bridging}.

\subsection{Ansatz for the Randomized Singular Value Decomposition}
\label{sec:rsvd}

Given that we chose an on-site Coulomb interaction, only particles of opposite spin interact. If we choose an ordering of variables in $\vec{n}$ such that all the spin-up variables $n_{ij\uparrow}$ appear in the first half, and spin-down in the second, we have a diffusion matrix with the structure
\begin{equation}
\label{eq:dmat}
\begin{split}
D = \begin{bmatrix} 
0 & D_q  \\
D_q^T  & 0
\end{bmatrix}
\end{split},
\end{equation}
see appendix~\ref{app:FPMatrices} for detailed expression of $D_q$. We define $D_{q,2} = -0.5iD_q$ and compute its Singular Value Decomposition (SVD), $D_{q,2} = USV^*$. Here, $S$ is a diagonal matrix with non-negative real entries and has simple positive square root. If we denote by $\overline{V}$ the (non-transposed) conjugate of $V$, we find that the new noise matrix $B$, defined as
\begin{equation}
\label{eq:bmat}
\begin{split}
B = \begin{bmatrix} 
U\sqrt{S} & iU\sqrt{S}  \\
i \overline{V}\sqrt{S}  & \overline{V} \sqrt{S} 
\end{bmatrix}
\end{split}.
\end{equation}
satisfies $D=BB^T$.

The analytical solution $B_0$, given in Appendix \ref{app:FPMatrices}, is a rectangular matrix of rank $4n_s$,  $B_0 \in \mathcal{M}_{2 n_s^2, 4 n_s}$, so the maximum rank of $D_q$ is $2n_s$.  
We can use this upper bound to improve the efficiency of the SVD by using a randomized algorithm (rSVD). 
We choose a rank $r$ and generate a random block matrix $Z \in \mathcal{M}_{n^2_s, r}$, calculate the product $D_qZ$, which serves as an approximation of the column space of $D_q$, then compute the SVD $D_q Z = USV$ on this reduced space. 
The result is subsequently projected back to the original space. 
The rank of the noise matrix $r$ could have been a good parameter to tune the method between accuracy and speed, but we will see in section \ref{sec:compTime} that the computations of the diffusion matrix calculation and of its decomposition with rSVD scale similarly. Hence, approximating the noise matrix with a rank $r$ smaller than the rank of diffusion matrix will give only marginal computational gains. 
We choose in the following examples a rank equal to the rank of $D_q$, half the diffusion matrix , for our Fermi-Hubbard model: $r=2n_s$. Then $D_qZ$ is not an approximation anymore, but compact writing of the full column space of $D_q$.
All details are in the original description of the algorithm \cite{halko2011finding}. For our numerical results, we have used the MATLAB implementation available on the MATLAB Central File Exchange \cite{rsvd2024}.

\section{Numerical Results}
\label{sec:num-exp}

\subsection{Practical Simulation Time}

Since the GPSR enables an exact mapping from the density operator to the probability distribution, the calculation of relevant quantities is limited only by sampling error, which can in principle be reduced by increasing the number of stochastic trajectories used for averaging. However, the spiking behavior arising in the numerical simulations, due to the spread of the probability density, is a practical limitation. We have shown that the practical simulation time can be extended by using the Takagi SVD method to compute the decomposition of $D$~\cite{rousse2023simulations}. Here, we show that the ansatz in Eqs.~\eqref{eq:dmat} and \eqref{eq:bmat}, combined with the use of the randomized SVD algorithm, can give more stable results while being substantially more efficient. In the following simulations, we measure the improvement in practical simulation time for systems of different sizes and geometry. The Hamiltonian is described in Eq.~\eqref{eq:FHeq0}, and the probability densities are sampled with $10^4$ stochastic trajectories.

\subsubsection{8-site system}

We first benchmark the efficiency of our method with a series of 8-site systems of different dimensions: an in-line 1D system, a 2D system of  $2\times 4$ sites, and a 3D system of $2 \times 2 \times 2$ sites.  The small size of these systems allows us to compare the GPSR and ED solutions and to explore the differences between 1D, 2D, and 3D systems. 

\begin{figure}[t]
\begin{center}
\subfigure[1D system.]{\includegraphics[height=4.4cm]{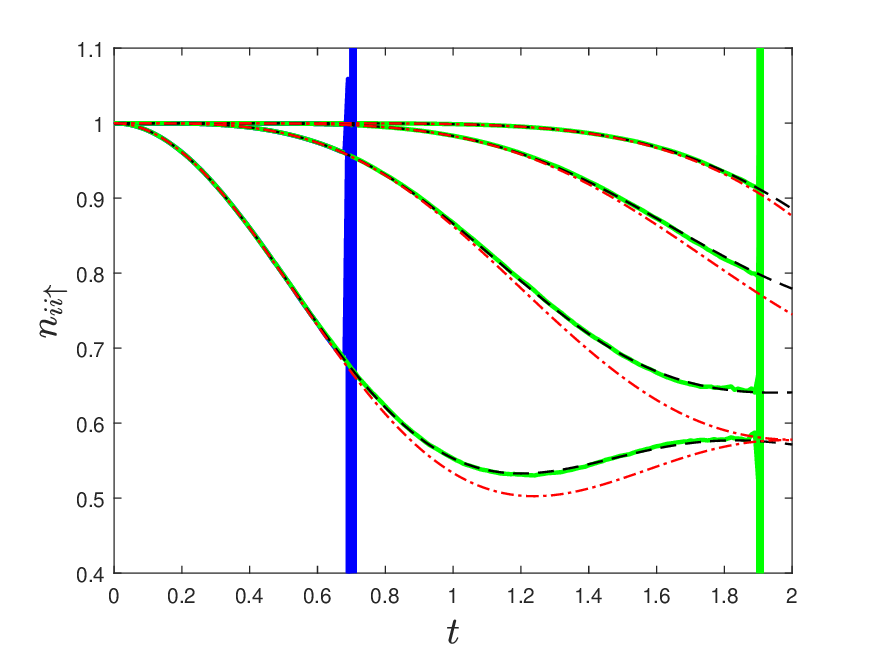}}
\hfill
\subfigure[2D system.]{\includegraphics[height=4.4cm]{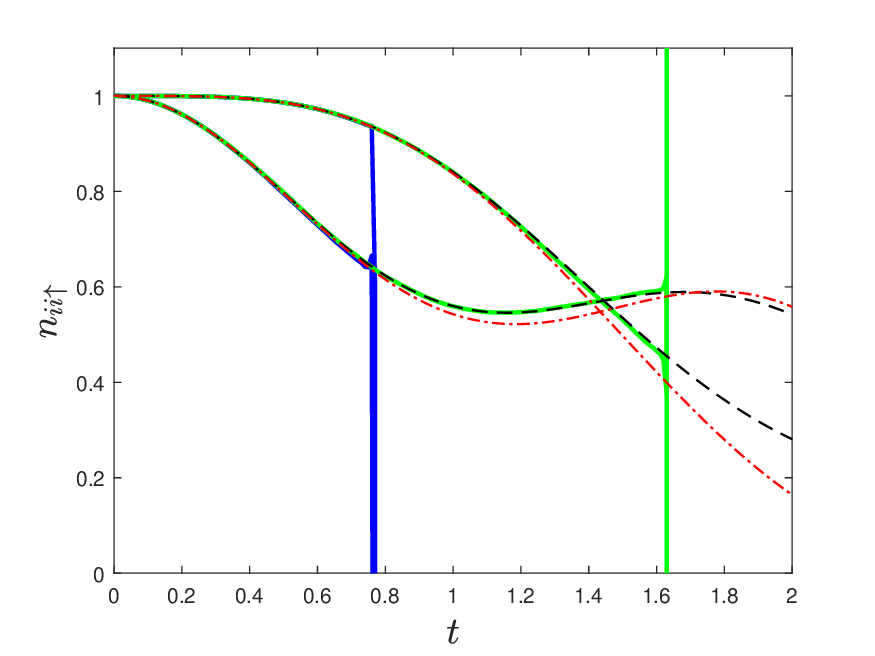}}
\hfill
\subfigure[3D system]{\includegraphics[height=4.4cm]{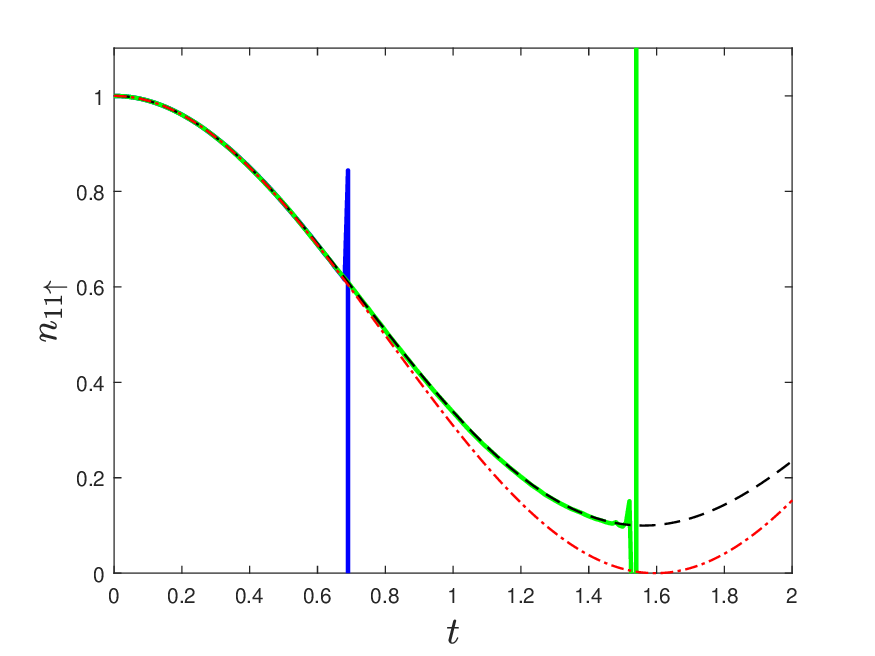}}
\caption{Site occupations of 8-site systems in different dimensions. The initial conditions are global spin waves, Eq.~(\ref{eq:IC8}), and we are presenting the spin-up occupation, $n_{ii \uparrow}$. The number of curves are different because symmetries cause some occupancies to be equal in the 2D and 3D systems. The black dashed curves represent the ED solutions, the red dash-dotted ones the HF solutions, the solid blue ones the GPSR solutions computed with the analytical noise matrix, and the solid green ones the GPSR solutions computed with the noise matrix from the randomized SVD decomposition. For all systems, switching from an analytical noise matrix (blue curves) to the noise matrix produced by the randomized SVD (green curves) at least doubles the practical simulation time of the GPSR solution.}
\label{fig:FH8_ni}
\end{center}
\end{figure}

We choose an initial occupation that is a ``global spin wave'', with particles of spin-up and the particles of spin-down starting on opposite sides of the system. We enumerate the sites with $i$ from $1$ to $8$, such that
\begin{equation}
    \label{eq:IC8}
    \begin{split}
    n_{ii\uparrow} = 1 \text{ and } n_{ii\downarrow} = 0  \text{, for } i \leq 4 ,\\
    n_{ii\uparrow} = 0 \text{ and } n_{ii\downarrow} = 1  \text{, for } i>4.
\end{split}
\end{equation}
In the 2D case, the spin wave is chosen such that there are opposite spins on each of the $2 \times 2$ sub squares. We plot in figure \ref{fig:FH8_ni} the spin-up site occupations of the first four sites, which are those starting with a spin-up fermion. The site occupations of the other sites are symmetric to an occupation already plotted. 

On the GPSR curves, we observe that the occupation loses continuity abruptly and stops. This phenomenon is what is termed  spiking: in a few time steps, one of the trajectories obtains values so large as to cause an overflow error, preventing calculation of the average from that point onwards. This limitated practical simulation time due to spiking is what we aim to increase. In each case, the GPSR method lasts at least twice as long with the numerical noise matrix in Eq.~(\ref{eq:bmat}), calculated from the decomposition of the diffusion matrix in Eq.~(\ref{eq:dmat}), as compared to the analytic noise matrix.  In particular, the extended simulation time means that the GPSR method is able to calculate quantities in the regimes where the Hartree-Fock approximation is inaccurate.

\subsubsection{Correlations}

One advantage of phase-space methods over the mean-field method is the handling of higher-order correlations. The correlation function $g^{(2)}$ gives the correlation of the occupation between two sites. The difference between the correlation computed with a mean-field method (HF) and with an exact one (ED or GPSR) gives us a better understanding of the entanglement between two particles. We define $g^{(2)}$ between sites $\lambda$ and $\kappa$ as
\begin{equation}
    \label{eq:g2}
g^{(2)}_{\lambda \kappa} = \frac{ \braket{ : \hat{n}_{\lambda \lambda}  \hat{n}_{\kappa \kappa} : }}{\braket{ : \hat{n}_{\lambda \lambda} :} \braket{:\hat{n}_{\kappa \kappa}:}} =  \frac{ \braket{\hat{c}^{\dagger}_{\lambda}  \hat{c}^{\dagger}_{\kappa} \hat{c}_{\kappa} \hat{c}_{\lambda} }}{\braket{ \hat{c}^{\dagger}_{\lambda} \hat{c}_{\lambda}} \braket{\hat{c}^{\dagger}_{\kappa} \hat{c}_{\kappa}}}
\end{equation}
where we have used ``$:.:$'' as notation for the normal ordering.

\begin{figure}[t]
\begin{center}
\subfigure[Simulation for $0 \le t \le 3$.]{\includegraphics[height=6.5 cm]{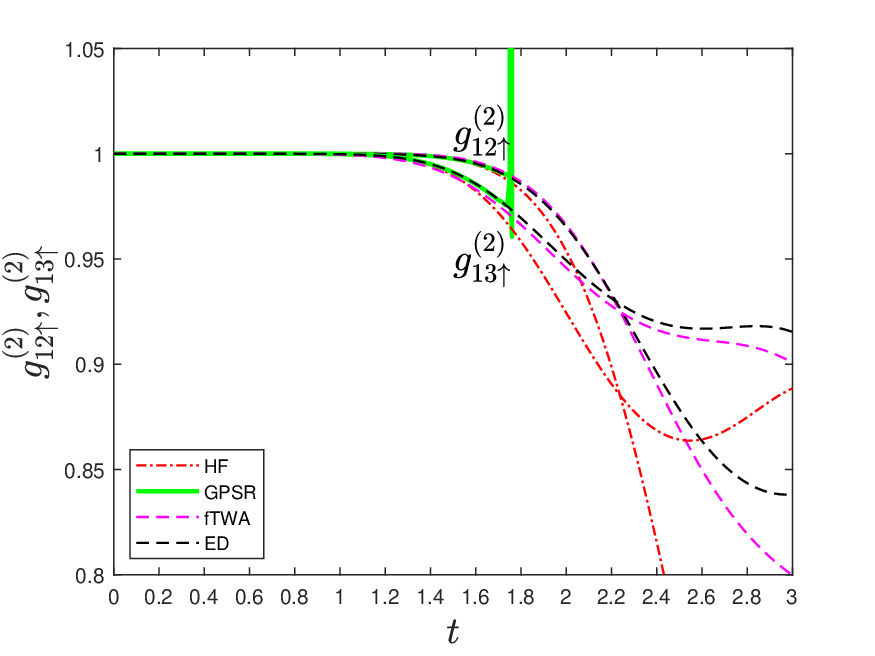}}
\subfigure[Zoom in for $t 1 \le t \le 2$.]{\includegraphics[height=6.5 cm]{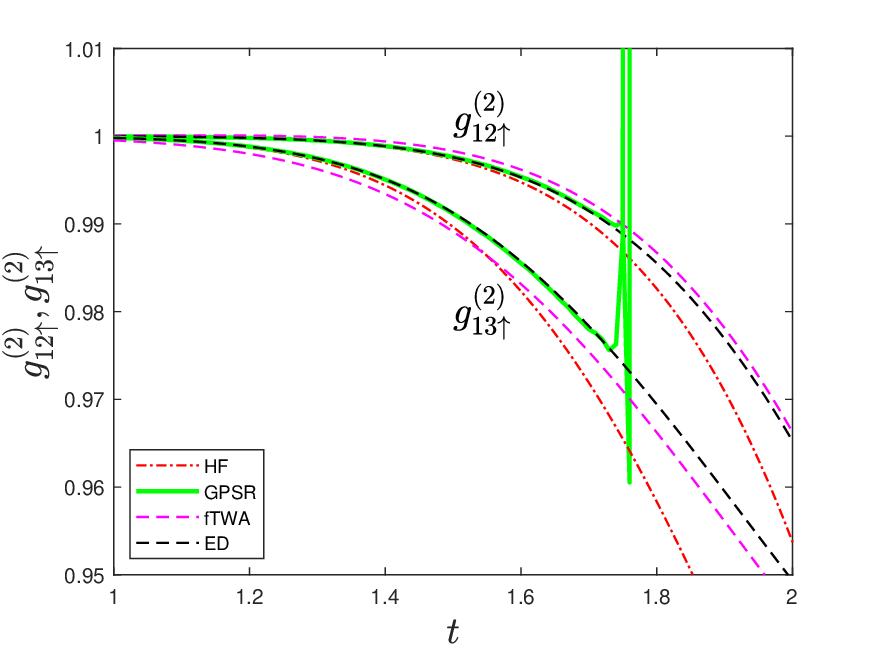}}
\caption{Pair-correlations between spin-up particles site occupations. For the 1D 8-site system, correlation between sites 1 and 2, then between sites 1 and 3, the next-to-nearest neighbor.}
\label{fig:FH8_g2_1}
\end{center}
\end{figure}

\begin{figure}
\begin{center}
\includegraphics[height=7 cm]{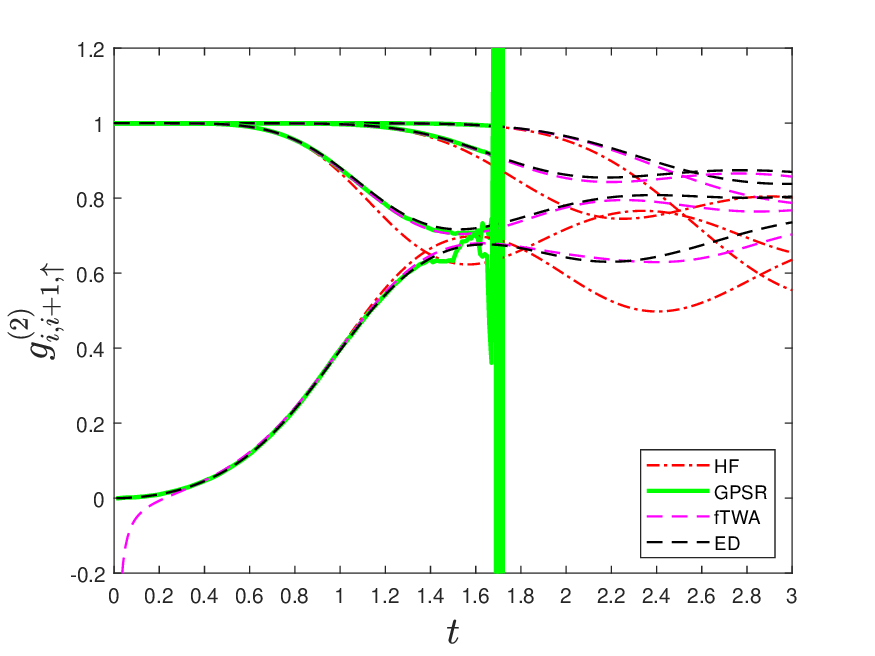}
\caption{Pair-correlations between spin-up particle sites occupation. For the 1D 8-sites system, correlation between sites 1 to 4 and their respective right neighbors (2 to 5). The curves starting at 0 are the correlation between sites 4 and 5 which start without spin-up particles, hence the initial difficulty of the fTWA to compute an accurate correlation. We have here used $10^4$ trajectories, but fTWA can still be improved visibly (closer to the ED solution) if we add up more trajectories.}
\label{fig:FH8_g2_2}
\end{center}
\end{figure}

\begin{figure}
\begin{center}
\includegraphics[height=7 cm]{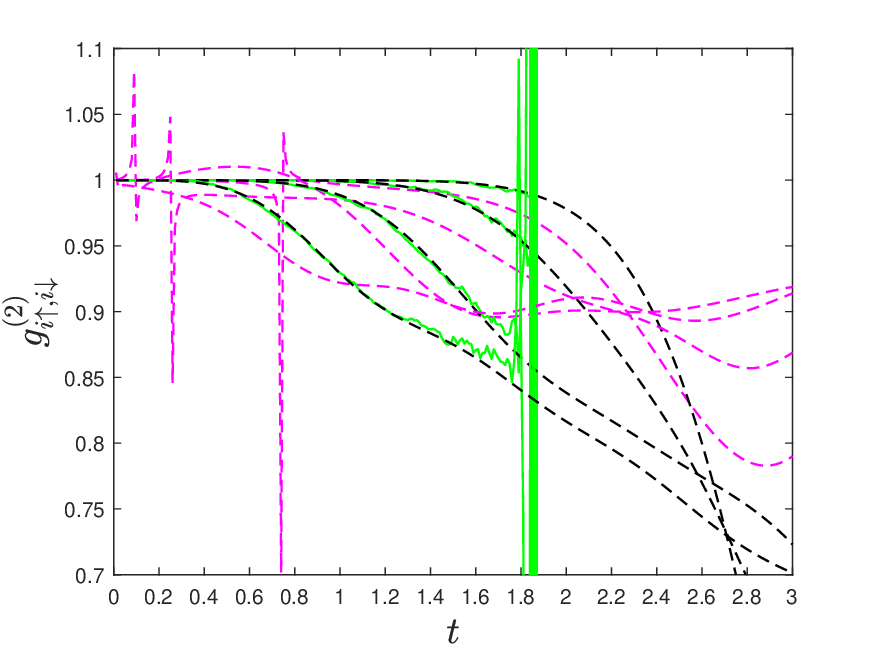}
\caption{On-site correlations on the 4 first sites. In the GPSR plots, the spikes announcing the failure of trajectories are clearer and appear sooner than in the occupation curves (Fig~\ref{fig:FH8_ni}). fTWA is still struggling to model correlation with near-to-zero occupation. We here have used $10^4$ trajectories, fTWA can still be visibly improved (closer to the ED solution) by adding up more trajectories. We did not plot the HF solution which is a constant 1 here.}
\label{fig:FH8_g2_0}
\end{center}
\end{figure}

In figures \ref{fig:FH8_g2_1}, \ref{fig:FH8_g2_2}, and \ref{fig:FH8_g2_0}, we have computed the second-order correlations between sites, in an 8-site linear system, for 4 methods (HF, GPSR, fTWA (see below), and ED). In figure \ref{fig:FH8_g2_1}, we plot the correlation between spin-up particles in sites 1 and 2 and in sites 1 and 3. The right panel zooms in on the most relevant time window. As expected, the correlation calculated with GPSR follows that of the ED solution. With the new numerical noise matrix, the practical simulation time goes beyond the point where HF and ED split away, so GPSR produces here new relevant information. We have also plotted with a magenta dashed line the solution of another phase-space method, the fermionic Truncated Wigner Approximation (fTWA) \cite{polkovnikov2010phase, rousse2023correlated}. The correlations calculated with fTWA are in this case accurate beyond the practical simulation time of GPSR.  In general, however, the method's accuracy is not guaranteed because the truncation in the method amounts to an uncontrolled approximation. Besides, $10^4$ trajectories do not appear to be enough for fTWA here, as we can see that more clearly in the next plots.

In figure~\ref{fig:FH8_g2_2}, we plot the correlation between spin-up particles in two neighboring sites. The curve starting from $g^{(2)}=0$ is $g^{(2)}_{4,5,\uparrow}$, the correlation between the last site filled with a spin-up particle and the first empty one. GPSR captures well the correlation from the start, whereas with fTWA poles appear. The fTWA variables are symmetrized so they are defined around zero, between $-0.5$ and $0.5$, which implies the denominator of the correlation, the double occupation, becomes null more frequently, hence the appearance of poles. Finally, figure \ref{fig:FH8_g2_0} plots the correlation between particles on the same site. The spikes preceding the end of the simulation in the GPSR curves are clear here. Once again, fTWA is giving incorrect results because of the near-to-zero denominators.

\subsubsection{36-site system}

\begin{figure}
\begin{center}
\includegraphics[height=6.5 cm]{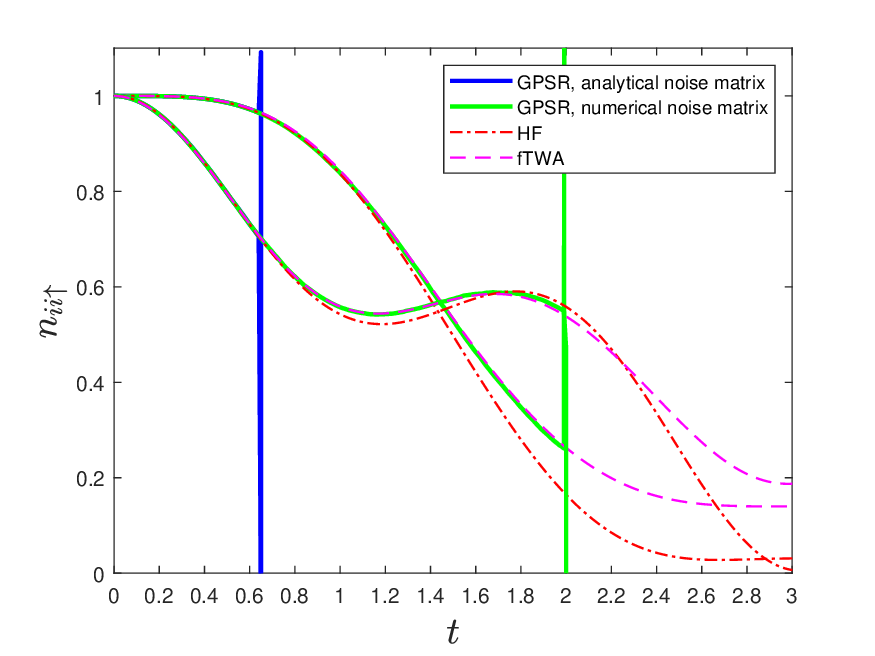}
\caption{Site occupations of a 3D 36-sites system ($4\times3\times3$). The initial condition is a global spin wave, and we are following the spin-up occupation of two sites. Dynamic simulation of systems of this size is unreachable with the ED method. On this system, the GPSR method with a numerical noise matrix (green line) has a practical simulation time three times larger than the one computed with the analytical noise matrix (blue line). We also observe that fTWA (magenta dashed line) produces an accurate estimation of the occupation until at least the moment GPSR fails.}
\label{fig:FH334_ni}
\end{center}
\end{figure}

As a demonstration of the favorable scaling of the GPSR with system system size, we now increase the lattice size to regimes that are out of reach of the ED method. We furthermore consider a 3D geometry, to contrast with the 1D situations where DMRG excels. We chose a 36-site, 3D system ($3 \times 3 \times 4$) with a spin wave as initial condition. We plot in figure~\ref{fig:FH334_ni} the occupation of sites 1, of coordinates $(1,1,1)$, and 10, of coordinates $(1,1,2)$; the other sites, by symmetry, share similar occupation dynamics. We observe that, again, the numerically determined noise matrix significantly extends the simulation time beyond that of the analytic noise matrix, into time-scales where HF gives discrepant results.

\subsection{Bell State}

With the GPSR method, we can also represent the states that are not accessible with the mean-field method, such as those with high entanglement. We model here a two-site Fermi--Hubbard model starting in a ``Bell state'', a superposition of the two Fock states where the two fermions are either on the first site or on the second one
\begin{equation}
\begin{split}
\ket{\Psi} = \alpha \ket{\uparrow \downarrow, 0} + \beta \ket{0,\uparrow \downarrow} 
\end{split},
\label{eq:BS}
\end{equation}
with $|\alpha|^2 + |\beta|^2 = 1$. We have chosen $\alpha = 1/\sqrt{2}$ and $\beta = i / \sqrt{2}$, because this dephasing produces occupancy changes. 

\begin{figure}
\begin{center}
\includegraphics[height=6.5 cm]{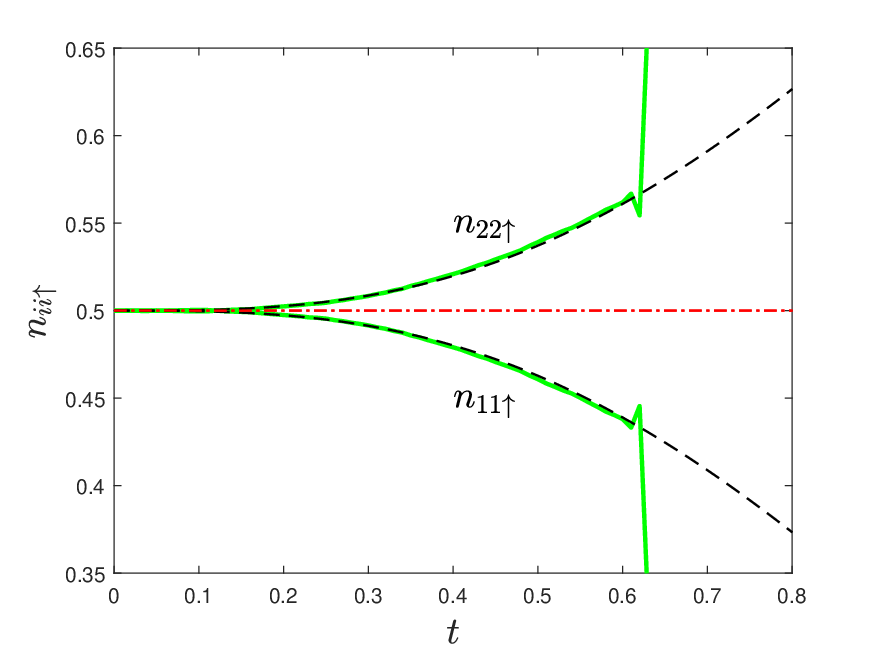}
\includegraphics[height=6.5 cm]{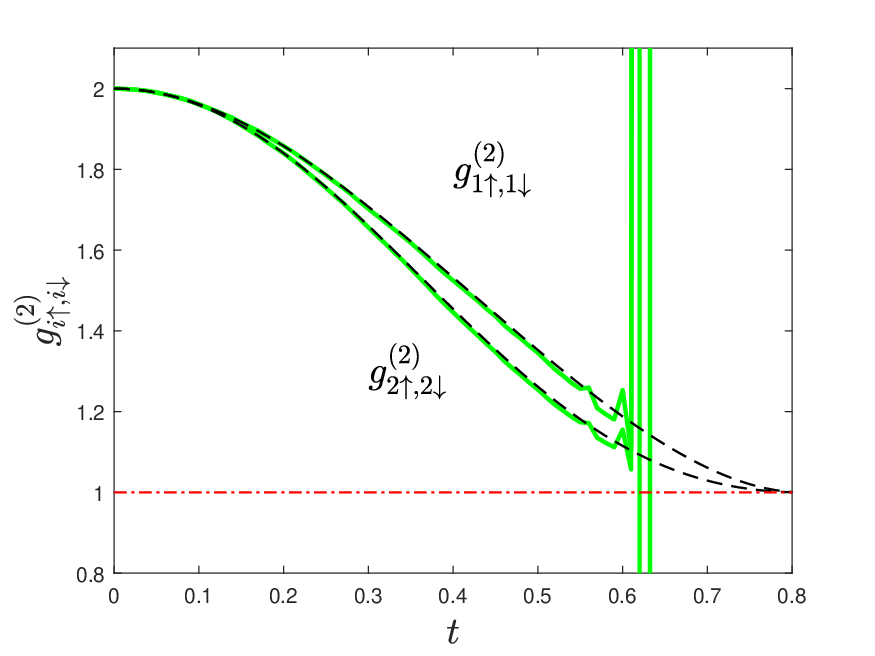}
\caption{Results for an entangled two-site system for the three different methods (GPSR, ED, HF). The colors are the same as in figure~\ref{fig:FH8_ni}. On the left, we see that the spin-up site occupations, on the right the normalized second-order correlation function $g^{(2)}$, between spin-up and spin-down particles on the same site.}

\label{fig:FH2_BS_nii}
\end{center}
\end{figure}

In figure~\ref{fig:FH2_BS_nii}, we follow the site occupations on the left, and the correlation function $g^{(2)}$ on the right. Those plots show clearly the ability that GPSR has to model a system that the HF approximation cannot handle. The Bell State is fully entangled, so the initial probability density of GPSR is no longer a Dirac delta function but has finite width. The corresponding noise already present at the beginning of the dynamic qualitatively shortens the practical simulation time, here $t_m \approx 0.6$, compared to figure~\ref{fig:FH8_ni}, for example. The probability density has also required more trajectories (here $10^5$) to reach the desirable accuracy.

\subsection{Complexity and computation time}
\label{sec:compTime}

The improvement of the practical simulation time comes at a substantial computational cost compared to using the analytical noise matrix. In the new method presented here, the computation of the diffusion matrix and its decomposition takes the majority of the simulation resources.

We first assess the computational cost of the new algorithm in terms of the number of sites $n_s$. 
For the calculation of the new noise matrix $B$, we decompose the diffusion matrix $D$ by computing the SVD of $D_q$ (see section \ref{sec:rsvd}). The computational cost of a classical SVD is dominated by the bi-diagonalization part, which requires $\mathcal{O}(nm^2)$ floating-point operations (flops) for a $n \times m$ matrix \cite{golub2013matrix}. Since $A \in \mathcal{M}_{n_s^2,n_s^2}$, the decomposition requires $\mathcal{O}(n_s^6)$ flops. When using the randomized SVD, the dominating part of the procedure is the orthonormalization of the image representation of $A$~\cite{halko2011finding}, which is a $n_s^2 \times n_s$ matrix. If this is performed using another SVD decomposition, the whole decomposition requires $\mathcal{O}(n_s^4)$ flops.

\begin{figure}
\begin{center}
\includegraphics[height=6.5 cm]{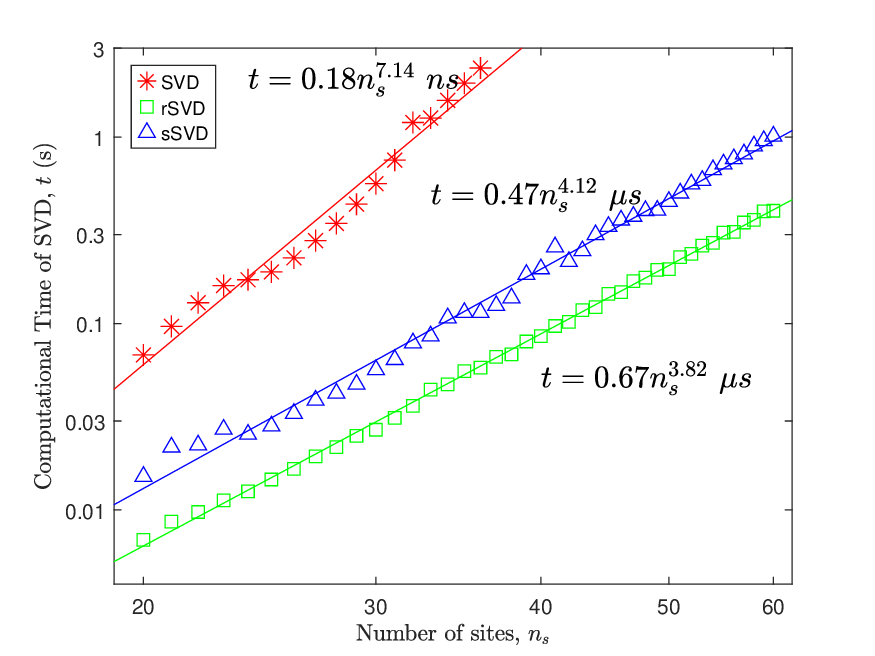}
\includegraphics[height=6.5 cm]{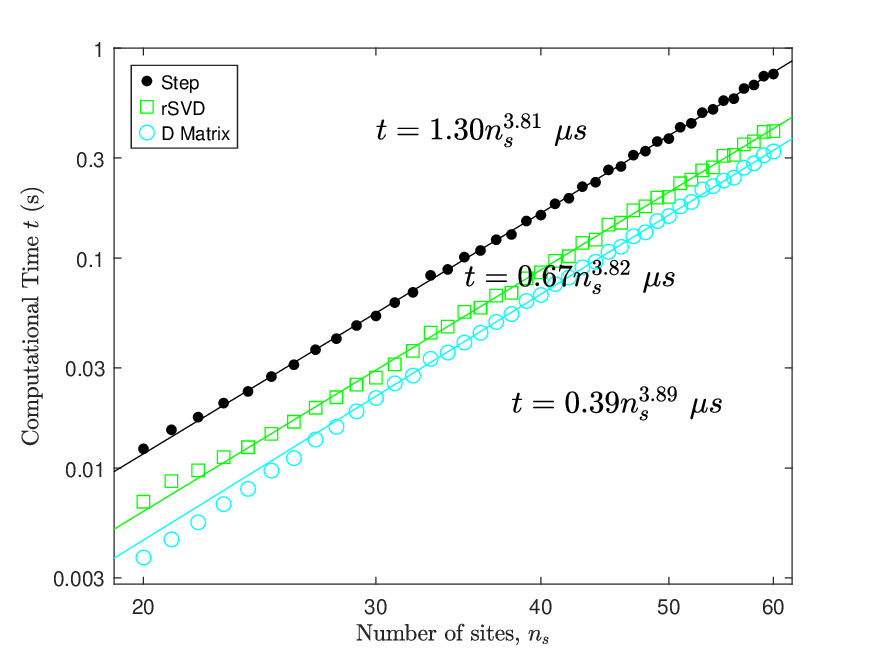}
\caption{Time scaling of the algorithm for the noise-matrix computation, log scale. On the left: mean timing of the decomposition of the diffusion matrix with a classical SVD in red, the low-rank SVD in blue, and the randomized SVD in green. On the right: mean computational time of a time step (in black) with its two main components: the computation of the diffusion matrix (in cyan), and its randomized SVD decomposition (in green). For each measure, we have computed and written the scaling parameter with a regression analysis, the solid curves are the scaling functions.}
\label{fig:compTimes}
\end{center}
\end{figure}

For a practical evaluation of the scaling, we measure the time taken to compute one trajectory depending on the system size. In figure~\ref{fig:compTimes}, we simulate the fermion dynamics of linear Fermi--Hubbard systems of increasing size to observe the increase of computational time needed. The simulations were done with 30 trajectories of 200 time point each. The left panel of figure~\ref{fig:compTimes} plots the computational time of the SVD decomposition with three different methods: the classical SVD method in red, a low rank SVD method \cite{yu2018efficient} (sSVD) in blue, and the randomized SVD one in green. We find time that the timing scales as $\mathcal{O}(n_s^{7.1})$ for the usual SVD, as $\mathcal{O}(n_s^{4.1})$ for the sSVD, and $\mathcal{O}(n_s^{3.8})$ for the randomized SVD. As expected, the computational time of a simulation is dominated by the matrix decomposition. In all our experiments, the randomized SVD is consistently the fastest one and gives trajectories of excellent stability.

In the right panel of figure~\ref{fig:compTimes} we have decomposed the noise-matrix computation with randomized SVD into its two components, the diffusion matrix computation in cyan, and its decomposition in green. The total computational time of a step, in black, roughly corresponds to the sum of these two parts. The two components scale similarly, as $\mathcal{O}(n_s^{\sim3.8})$, so improving only the decomposition efficiency of $D$ further, for example by reducing the number of singular values kept in the decomposition, would not bring a significant time gain. 

The practical computational times take into account much more than the number of multiplications, so the correspondences between the computational time and the actual timings are, as expected, far from being exact. We should not forget that they depend on the coding language used, here MATLAB, on the features of the machine used to run the experiments, and on the fine-tuning of the implementations.

\section{Summary and outlook}
\label{sec:conclusion}

Although quantum phase-space methods can be used to simulate many-body systems of sizes far beyond the scope of exact diagonalisation, they are in practice limited to short-time dynamics.  It is this key limitation that we have addressed in this paper.  Our previous attempt to increase this practical simulation time \cite{rousse2023simulations} by computing the noise matrix with the Takagi decomposition faced the instability of the decomposition for large systems. Here, we use the randomized SVD, which has no problem of instability and still more than doubles the practical simulation time.

We showed the efficiency of the methods on small systems of different dimensionality, on a very entangled system of two sites, and on a large 3D-system. We have also estimated the computational cost of the method, which requires $O(n^4)$ flops. An exact many-body dynamic on a 3D, 60-sites Fermi--Hubbard system is currently out of reach for ED or DMRG. The timescale of the GPSR is still limited, but, with the numerically determined noise matrix, it extends approximately twice as long as previous implementations. 

A typical value of the tunneling interaction $J$ for metals would be $J=1eV$, which corresponds to time units of $0.6 fs$~\cite{Will2013}. With our current method and a Coulomb interaction of $U=1$, we can then simulate exactly the electron dynamic in a metallic sample for $1 fs$, which is still too short for most relevant physical phenomena. The practical simulation time also depends greatly on the value of $U$: a lower value would increase the practical simulation time, whereas a greater one would decrease it. We chose this intermediate value of $U=J=1$ because it corresponds to the beginning of the strong correlation regime.  

From this point of view, the next steps should focus on further improving the practical simulation time. We hope that understanding why the numerical noise-matrices work better than the analytical ones will help us find better decompositions of the diffusion matrix. It may also be possible to find physically relevant simulation that would suit the GPSR strengths and weaknesses: large 2D or 3D system, of the order of 50-100 sites, and short simulation time span. Although it may need a demonstration of its value on a physical applications, it is a promising method that can already be used to evaluate the accuracy of other methods on models too large for ED. 

Although we have focused on the GPSR and the Hubbard model, the technique of numerical diffusion gauges demonstrated here could also be applied to other physical models and indeed other representations within the broad class of phase-space methods.

\section{Acknowledgments} 

F.R. and M.\"{O}. are supported by Carl Tryggers foundation and RR-ORU-2023/2024. F.R. acknowledges support from the Swedish Research Council, the Knut and Alice Wallenberg Foundation, the European Research Council via Synergy Grant No. 854843- FASTCORR, Swedish energy agency and eSSENCE. The work of A.D. was supported by the Swedish Research Council (VR) under grant 2021-05393.

\bibliography{bibliography_GPSR.bib}
\bibliographystyle{apsrev4-1}

\appendix

\section{Derivation of the SDE}
\label{app:FPMatrices}

For the derivation of the stochastic differential equations, we denote in Greek letters the pairs site-spin $\alpha = (a,\sigma_a)$; $\vec{n}$ denotes the phase-space variables, and $(\vec{n})_{\lambda \kappa} = n_{\lambda \kappa} $ denotes the whole one-particle density matrix in GPSR, Using the mapping relations of Eqs.~\eqref{eq:mapping}, we find the Fokker--Planck equation in its general form
\begin{equation}
\label{eq:FK1}
\frac{\partial }{\partial t} P(\vec{n},t) = \left[ -\sum_{\lambda \kappa} \frac{\partial}{\partial n_{\lambda \kappa}} A_{\lambda \kappa}(\vec{n}) + \frac{1}{2} \sum_{\lambda \kappa, \mu \nu} \frac{\partial}{\partial n_{\lambda \kappa}} \frac{\partial}{\partial n_{\mu \nu}}D_{\lambda \kappa, \mu \nu}(\vec{n}) \right]P(\vec{n},t).
\end{equation} 
where for the drift matrix $A(\vec{n})$ we have
\begin{equation}
\label{eq:drift1}
A_{\lambda \kappa} (\vec{n}) = \frac{i}{\hbar}  \sum_{\alpha} \left(  J_{\alpha \lambda} n_{\alpha \kappa} - J_{\kappa \alpha} n_{\lambda \alpha} \right) ,
\end{equation}
and for the diffusion matrix $D(\vec{n})$ we have
\begin{equation}
\label{eq:diff1}
D_{\lambda \kappa, \mu \nu} (\vec{n}) = 2i \sum_{\alpha \beta} U_{\alpha \beta} ( \tilde{n}_{\alpha \kappa} n_{\lambda \alpha}  \tilde{n}_{\beta \nu} n_{\mu \beta} -  n_{\beta \kappa} \tilde{n}_{\lambda \beta} n_{\alpha \nu} \tilde{n}_{\mu \alpha} ) .
\end{equation}

We recall that $\tilde{n}_{\lambda \kappa} = \delta_{\lambda \kappa} - n_{\lambda \kappa}$ is the hole  number. Now, as we work at constant particle number and in the absence of spin-flip terms, we only need to consider pairs of indices of the same spin: $n_{\lambda \kappa}$ becomes $n_{ij\sigma}$. We also focus on on-site constant interactions:  $U_{a\sigma_a, b \sigma_b} = \frac{U}{2} \delta_{ab} \delta_{\sigma_a, -\sigma_b}$, so we can write Eqs.~(\ref{eq:drift1}) and (\ref{eq:diff1}) as
\begin{equation}
\label{eq:drift2}
A_{ij\sigma} (\vec{n}) = \frac{i}{\hbar}  \sum_{a} \left(  J_{a i} n_{a j \sigma} - J_{j a} n_{i a \sigma} \right) ,
\end{equation}
\begin{equation}
\label{eq:drift2}
D_{ij \sigma, kl \sigma'}   = i U \delta_{\sigma, -\sigma'} \sum_{a} ( \tilde{n}_{a  j \sigma} n_{i a\sigma}  \tilde{n}_{a  l \sigma'} n_{k a \sigma'} -  n_{a  j \sigma} \tilde{n}_{i a \sigma} n_{a  l \sigma'} \tilde{n}_{k a \sigma'}  ) .
\end{equation}

To turn the Fokker--Planck equation into a SDE, we have to compute a noise matrix $B$ such as $BB^T = D$. For the diffusion matrix of the Fermi--Hubbard model with on-site interactions, we have an analytical solution that we call $B_0$
\begin{equation}
\label{eq:B0Matrix}
\begin{split}
B_0 = \begin{bmatrix} 
B^{(1)}_{\uparrow} & i B^{(1)}_{\uparrow} & B^{(2)}_{\uparrow} & i B^{(2)}_{\uparrow} \\
B^{(1)}_{\downarrow} & -i B^{(1)}_{\downarrow} & B^{(2)}_{\downarrow} & -i B^{(2)}_{\downarrow}
\end{bmatrix},
\end{split}
\end{equation}
with $B^{(1)}_{\sigma}$ and $B^{(2)}_{\sigma}$ matrices of size $(n_s^2 \times n_s)$ with elements
\begin{itemize}
    \item $(B^{(1)}_{\sigma})_{ij,p} = \frac{\sqrt{ i u}}{\sqrt{2}} n_{pi\sigma} \tilde{n}_{jp\sigma}$,
    \item $(B^{(2)}_{\sigma})_{ij,p} =  i \frac{\sqrt{ i u}}{\sqrt{2}}  \tilde{n}_{pi\sigma} n_{jp\sigma}$.
\end{itemize}

We can either directly use this noise matrix to calculate trajectories, or we can compute the corresponding diffusion matrix
\begin{equation}
\begin{split}
D = B_0 B_0^T= 2 \begin{bmatrix} 
0 &  B^{(1)}_{\uparrow} B^{(1),T}_{\downarrow} + B^{(2)}_{\uparrow} B^{(2),T}_{\downarrow}  \\
 B^{(1)}_{\downarrow} B^{(1),T}_{\uparrow} + B^{(2)}_{\downarrow} B^{(2),T}_{\uparrow}  & 0
\end{bmatrix},
\end{split}
\end{equation}
and decompose it to find another noise matrix.

\section{Study of individual trajectories}
\label{app:indTraj}

\begin{figure}[h!]
\begin{center}
\includegraphics[height=8 cm]{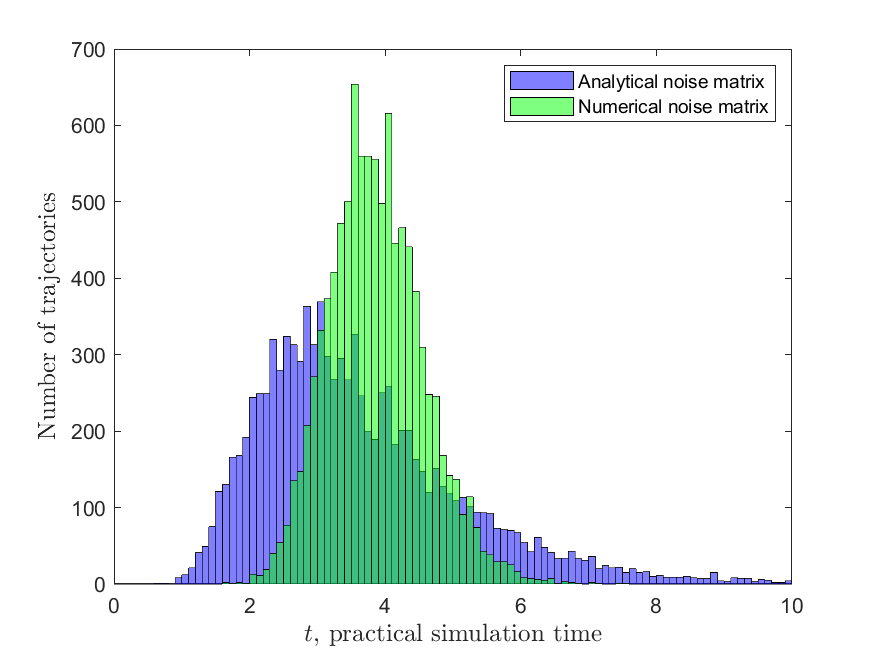}
\caption{Practical simulation time of individual trajectories. In blue, trajectories are computed with the analytical noise matrix $B_0$, in green with the noise-matrix computed with a random-SVD of $D$. Using numerical noise-matrices does not improve the average practical simulation time, but it reduces its variance so it increases the probable minimum.  }
\label{fig:practTimes1}
\end{center}
\end{figure}

In figure \ref{fig:practTimes1}, we observe that using the numerical version of the noise matrix $B$ is not a global improvement for all trajectories, but the average practical simulation time of trajectory validity is roughly conserved, while its variance is reduced. This leads to an increase in the practical simulation time of the worst trajectory and delays the onset of the spikes.

We must confess that we do not understand in detail why computing the noise matrix from the ansatz \eqref{eq:dmat} and \eqref{eq:bmat} leads to a longer practical simulation time. We have noticed that the numerical noise matrix has a reduced rank compared with the analytical one. $B_0$ as computed with Eq.~\eqref{eq:B0Matrix} is a $2 n_s^2 \times 4n_s$ matrix of rank $4n_s-2$, but $D$ and hence $B$ are of rank $4n_s-4$. We cannot find a noise matrix of lower rank than the diffusion matrix itself, and if the rank reduction is the only reason for the improvement observed, we may have reached the maximum gain possible with diffusion gauge.

\section{Initial condition of a Bell state for GPSR} \label{app:bs_gpsr}

A Bell state, as described in Eq.~\eqref{eq:BS}, is not a Slater determinant, so its corresponding phase-space probability density $P(\vec{n})$ is not a Dirac delta function. \cite{corney2005gaussian, corney2006gaussian}. The only non-zero number observables of the Bell state are
\begin{itemize}
    \item $\bra{\Psi} \hat{n}_{11\sigma} \ket{\Psi} = |\alpha|^2$ 
    \item $\bra{\Psi} \hat{n}_{22\sigma} \ket{\Psi} = |\beta|^2$
    \item $\bra{\Psi} \hat{n}_{11\uparrow} \hat{n}_{11\downarrow} \ket{\Psi} = |\alpha|^2$ 
    \item $\bra{\Psi} \hat{n}_{22\uparrow} \hat{n}_{22\downarrow} \ket{\Psi} = |\beta|^2$ 
    \item $\bra{\Psi} \hat{n}_{12\uparrow} \hat{n}_{12\downarrow} \ket{\Psi} = \alpha^*\beta$ 
    \item $\bra{\Psi} \hat{n}_{21\uparrow} \hat{n}_{21\downarrow} \ket{\Psi} = \alpha \beta^*$ 
\end{itemize}
    
All the other first- and second-order observables are null, notably $\bra{\Psi} \hat{n}_{11\sigma} \hat{n}_{22\sigma'} \ket{\Psi}=0$, which denotes negative correlations. From those observables, we build a probability density $P(\vec{n})$ with the proper moments. Using the Wick theorem, Eq.~\eqref{eq:WickTh}, we can compute the moments of $P(\vec{n})$ and choose a corresponding stochastic initial condition. Let $W_1$, $W_2$, and $W_3$ be independent real normal random variables, and let us define
\begin{equation}
\begin{split}
    \begin{bmatrix}
    n_{11\uparrow}  &  n_{12\uparrow} \\ 
    n_{21\uparrow}  &  n_{22\uparrow}  \\
    \end{bmatrix} &=
    \begin{bmatrix}
    |\alpha|^2  & 0  \\ 
    0  & |\beta|^2  \\
    \end{bmatrix} + 
    \begin{bmatrix}
    |\alpha|^2 W_1  & \alpha^* W_2  \\ 
    \beta^* W_3  & -|\beta|^2 W_1  \\
    \end{bmatrix}, \\
    \begin{bmatrix}
     n_{11\downarrow} & n_{12\downarrow}\\
     n_{21\downarrow} & n_{22\downarrow}
    \end{bmatrix} &=
    \begin{bmatrix}
        |\alpha|^2  &  0  \\ 
        0  & |\beta|^2  \\
    \end{bmatrix} +
    \begin{bmatrix}
    |\alpha|^2 W_1  & \beta W_2  \\ 
    \alpha W_3  & - |\beta|^2 W_1  \\
    \end{bmatrix}.
    \end{split}
\end{equation}
Each trajectory starts with an independent realization of this density matrix. We have experimented with two stochastic variables, Gaussian noise, $W_i \sim \mathcal{N}(0,1)$ and a binary stochastic variable, $\Omega(W_i) = \{ -1,1 \}$. The binary variable, with a sample space $\Omega$ being as close to zero as possible, produces the best practical simulation time, so we used it for the dynamics in figures~\ref{fig:FH2_BS_nii}. This representation of the Bell state produces a probability density composed of eight distinct points in the real-valued phase-space.

\end{document}